\documentclass[pra,aps,amsmath,amssymb,superscriptaddress,longbibliography,notitlepage,twocolumn]{revtex4-1}
\usepackage{graphicx,multirow,outlines,ulem}
\usepackage{color}
\usepackage{hhline}
\usepackage{physics}
\usepackage{hyperref}

\usepackage[sc,osf]{mathpazo}

\usepackage[mathscr]{eucal}

\newcommand{\bla}{\color{black}}

\newcommand{\huk}{\hookrightarrow}

\usepackage{tikz}
\usepackage{xcolor}

\newcommand{\bo}{\begin{outline}}
\newcommand{\eo}{\end{outline}}

\newcommand{\qed}{\nobreak \ifvmode \relax \else
      \ifdim\lastskip<1.5em \hskip-\lastskip
      \hskip1.5em plus0em minus0.5em \fi \nobreak
      \vrule height0.75em width0.5em depth0.25em\fi}
\begin{document}

\title{Observational-entropic study of Anderson localization} 
\author{Ranjan Modak}
\email{ranjan@iittp.ac.in}    
\author{S. Aravinda} 
\email{aravinda@iittp.ac.in}
\affiliation{ Department of Physics, Indian Institute of Technology Tirupati, Tirupati, India~517619} 
\begin{abstract}
The notion of the thermodynamic entropy in the context of quantum mechanics is a controversial topic. While there were proposals to refer von Neumann entropy as the thermodynamic entropy, it has it's own limitations. 
The observational entropy has been developed as a generalization of Boltzmann entropy, and it is presently one of the most promising candidates to provide a clear and well-defined understanding of the thermodynamic entropy in quantum mechanics. In this work, 
we study the behaviour of the observational entropy in the context of localization-delocalization transition for one-dimensional Aubrey-Andr\'e (AA) model. We find that for the typical mid-spectrum states, in the delocalized phase the observation entropy grows rapidly with coarse-grain size and saturates to the maximal value, while in the localized phase the growth is logarithmic. Moreover, for a given coarse-graining, it increases logarithmically with system size in the delocalized phase, and obeys area law in the localized phase. We also find the increase of the observational entropy followed by the quantum quench, is logarithmic  in time in the delocalized phase as well as at the transition point, while in the localized phase it oscillates. Finally, we also venture the self-dual property of the AA model using momentum space coarse-graining.

\end{abstract}

\maketitle

\section{Introduction} 
Thermodynamic entropy within quantum mechanics (QM) is a debatable topic, and von Neumann, himself was not comfortable  to define von Neumann entropy as thermodynamic entropy \cite{von2010proof,She99,HS06,MMB05,DRE+11,AY13}. The problem is more evident and explicit in the studies of foundational questions in quantum mechanics \cite{Per89,Weinberg89Pre,Wei89,HW13}. Motivated by the classical Boltzmann entropy, Safranek, Deutch, and Aguire generalized it to quantum systems (see for review \cite{vsafranek2021brief}), and they call it observational entropy \cite{deutchquantum2019, quantum2020deautch}. They show that the observational entropy is a monotonic function of the coarse-graining. 
Classical variant of observational entropy has been discussed in the early days of classical and quantum statistical mechanics \cite{Caves_notes,gibbs1902elementary,ehrenfest1990conceptual,von2010proof,strasberg2021second}. 
In recent days, the comparison of observational entropy with the entanglement entropy \cite{faiez2020typical}, and with the classical definition of observational entropy, \cite{vsafranek2020classical}, information-theoretic extensions using Petz recovery formalism \cite{Buscemi_22}, as a measure of correlations \cite{schindler2020quantum}, witnessing quantum chaos \cite{pg2022witnessing}, and using it to derive microscopically the laws of thermodynamics \cite{strasberg2021second} are few examples which have brought lots of attention. Specially, given that the entanglement entropy, played an important role in studying many-body properties , and for the larger system size as thermodynamic entropy \cite{deutsch2010thermodynamic,deutsch2013microscopic,santos2012weak}, the comparison  between observational entropy and entanglement entropy become even more crucial in the context of many-body quantum systems.


On the other hand, quantum computational  simulation \cite{georgescu2014quantum} of many-body systems would be a test bed for all our future technological ventures and also for fundamental understandings \cite{altman2021quantum}. The main issue in the quantum computer (QC) is that the output of the QC will be a probability distribution of certain measurements. The quantification of any properties should be a function of outcome statistics in order to proclaim the truly quantum behaviour. 
Observational entropy, which is a directly measurable quantity, will be a diagnostic tool to investigate the many-body systems in quantum computational simulations. 

In this work our main goal is to investigate whether observational entropy can be used as a diagnostic tool to detect localization-delocalization transition.  
\textcolor{black}{There has been a plethora of theoretical work \cite{abanin.2016,iyer.2013,lev1,lev2,lev3,vosk.2013,moore.2012,rmp_localization} which uses the scaling of the entanglement entropy with the system size (it obeys volume law in the delocalized phase and area law in the localized phase) as a probe to detect the transition. 
But, the entanglement entropy is not a directly measurable quantity \cite{islam2015measuring}. On the other hand, the fact that observational entropy is a directly measurable  quantity and can easily be measured in the current experimental setups \cite{carr2009cold,safronova2018search,atature2018material,vandersypen2019semiconductor,hensgens2017quantum,
hartmann2008quantum,vaidya2018tunable,norcia2018cavity,davis2019photon,wang2015topological,noh2016quantum,
hartmann2016quantum,kjaergaard2020superconducting,ganzhorn2019gate,kandala2017hardware,hempel2018quantum,nam2020ground,britton2012engineered,bohnet2016quantum}, makes our study even more relevant.}
To investigate the localization-delocalization transition, the best suited Hamiltonians is
the Aubry-André
(AA) Hamiltonian \cite{aubry.1980}.  While in one dimension, any arbitrary weak amount of true disorder is
sufficient to localize all eigenstates of a noninteracting system (which is famously known as Anderson localization\cite{anderson.1958,tvr.1979,tvr.1985}), 
this model has the incommensurate on-site potential, and
the localization-delocalization transition occurs for a finite
incommensurate potential amplitude. Also, in contrast to the true disordered models, the model with incommensurate on-site potentials can be successfully realized in the ultra-cold atom experiments~\cite{rmp_localization,kaufman2016quantum}.

\section{Observational entropy \label{sec:obs_ent}} 
Here we we define observational entropy and list some of its properties which will be useful for our analysis. 
Consider the Hilbert space $\mathcal{H}$ of dimension $d$. Let $\Pi_i$ be the set of projection operators with the completeness properties $\sum_i \Pi_i = I$, which  represents the measurement.  The projectors $\Pi_i$ projects the state vectors into orthogonal subspaces $\mathcal{H}_i$ and the total Hilbert space is partitioned as $\mathcal{H} = \bigoplus_i \mathcal{H}_i$. Each of the subspace $\mathcal{H}_i$ can be treated as a macrostate and the  probability of finding the state of the system $\rho$ in the subspace $\mathcal{H}_i$ (macrostate) is given as  $p_i = \tr(\Pi_i \rho)$. Hence the volume of the macrostate is simply the volume of the subspace $\mathcal{H}_i, ~V_i = \tr(\Pi_i)$.  

The coarse-graining $\chi$  is specified by the set of  projectors $\Pi_i$ with $\sum_i \Pi_i = I$. For any two coarse-graining $\chi_1$ and $\chi_2$ with projector sets $\{\Pi_{i_1}\}$ and $\{\Pi_{i_2}\}$, $\chi_1$ is said to be {\it rougher}  than $\chi_2$, and is represented as $\chi_1 \huk \chi_2 $,  if there exists an index set $N_{(j_1)}$ such that 
\begin{equation*}
    \Pi_{i_1} = \sum_{i_2 \in N_{i_1}} \Pi_{i_2} \quad \forall \Pi_{i_1}.  
\end{equation*}
$\chi_1 \huk \chi_2 $ can also be read as $\chi_2$ is \textit{finer} than $\chi_1$.  

The observational entropy of the system in the state $\rho$ with the coarse-graining $\chi$ is defined as follows \footnote{Since we are concerned on studying observational entropy, unless otherwise specified, $S$ denotes observational entropy and we don't use subscript/superscript to denote observational entropy. Other type of entropies are denoted with suitable subscripts}: 
\begin{equation}
    S_\chi (\rho) = -\sum_i p_i \ln p_i + \sum_i p_i \ln V_i
    \label{eq:Ob_ent} 
\end{equation}

 The observation entropy $S_\chi$  is a monotonic function of the coarse-graining. If  $\chi_1$ is rougher than $ \chi_2 $, $\chi_1 \huk \chi_2$, then 
\begin{equation}
    S_{\chi_1} (\rho) \geq S_{\chi_2} (\rho) 
    \label{eq:cg_def} 
\end{equation}
The coarse-graining $\chi$ specified with the single element identity $I$, which is rougher than the all coarse-grainings, i.e., $\chi_I \huk \chi ~ \forall \chi$. The observational entropy is maximum for the coarse-graining $\chi_I$, $S_{\chi_I} = \ln d $ and  can be termed as roughest coarse-graining. Another extreme of the coarse-graining is the finest coarse-graining for which the coarse-graining conatins all rank-1 projectors, i.e., $\chi_{fn}$ with $\Pi_i$ such that $V_i = 1 ~ \forall i$.  The coarse-graining relation $\huk$ is a partial order hence for any other coarse-graining $\chi$, the observational entropy will be 
\begin{equation}
    S_{\chi_{fn}} (\rho)  \leq S_\chi (\rho) \leq S_{\chi_I} (\rho). 
    \label{eq:fineC} 
\end{equation}

\bla 
\section{Model}
We study a system of fermions in an one-dimensional lattice of size $L$, which is described by  the following  Hamiltonian:
\begin{eqnarray}
{H}&=&-\sum_{j=1}^{L-1}({c}^{\dag}_j{c}_{j+1}+\text{H.c.})+\Delta\sum_{j=1}^{L} \cos(2\pi\alpha j+\phi){n}_j, \nonumber \\
\label{nonint_model}
\end{eqnarray}
where ${c}^{\dag}_j$  ( ${c}_j$) is the fermionic creation (annihilation) operator at site $j$, ${n}_j ={c}^{\dag}_j{c}_{j}$ is the number operator, and  $\alpha$ is an irrational number. Without loss of any generality, we choose $\alpha=\frac{\sqrt{5}-1}{2}$, and $\phi$ is a random number chosen between $[0,2\pi]$. We do averaging over $\phi$ for all the calculations presented in this work to obtain  better statistics. 
The Hamiltonian ${H}$ is known as Aubry-Andr{\'e} (AA) model. Unlike Anderson model, this model  supports a  delocalization-localization transition as one tunes $\Delta$ even in 1 dimension (1D). In the thermodynamic limit, $\Delta=2$ corresponds to the transition point \cite{aubry.1980}  between localized and delocalized phases and  for $\Delta<2$ ($\Delta>2$), all the eigenstates of the model are delocalized (localized). Another very interesting property of the AA model is its self-duality.  
Upon Fourier transformation, which exchanges real and momentum space, AA model can be shown to be dual to itself, with $\Delta/2 \to (\Delta/2) ^{-1}$ ~\cite{aubry.1980}. Since
the eigenstates of this model are localized in real space
for large $\Delta$, they are localized in momentum space for
small $\Delta$. Furthermore, given in AA model, there is only a single transition between localized and delocalized phases, it must be at exactly $\Delta/2 = 1$, where the model is invariant under the duality.


Given the Hamiltonian is defined over a lattice 
of length $L$, the finest graining implies a scenario having $L$ projectors  each of them defined over an individual lattice site i.e.  $\{\Pi_{x _1}=|x\rangle\langle x|\}$ and on the other hand the roughest coarse-graining implies only a single projector $\Pi_{x_L}=\sum_{x=1}^{L} |x\rangle\langle x|$ which incorporate all the lattice sites. Note that $V_{x _1}=1$ and $V_{x_L}=L$ for the finest coarse-graining and the roughest coarse-graining respectively. In order to investigate the effect of other possible coarse-graining, we divide our lattice into $L/m$ equal parts, and where $m=1$, 2, 4, 8, 16, $\cdot \cdot 2^{\ln_2L}$, and $L$ is chosen such a way that $\ln_2L$ is an integer. It implies that the coarse-graining length $V_{x_m}=m$ and the corresponding projectors are  $\{\Pi_{x_m}=\sum^{m\eta}_{x=m(\eta-1)+1} |x\rangle \langle x|\}$, 
where $\eta=1$, $2, \cdot \cdot L/m$. Note that for a given coarse-graining $\chi \equiv  x_m$, $\sum_{x_m} {\Pi}_{x_m}=I$
and ${\Pi}_{x_m}{\Pi}_{x'_m}=\delta_{x_m x'_m}I$, the observational entropy corresponds to state ${\rho}$ from the definition (\ref{eq:Ob_ent}), is
\begin{equation}
S_{\chi}({\rho})=-\sum_{x_m} p_{x_m}\ln p_{x_m}+p_{x_m}\ln V_{x_m}, 
\end{equation}
where $p_{x_m}=\Tr[{\rho}{\Pi}_{x_m}]$ and $V_{x_m}=\Tr[{\Pi}_{x_m}]$.

Given that the notion of the observation entropy is not restricted to the real space coarse-graining, one can define any observable, it's projectors 
can be used to define the coarse-graining. At the later part of the paper, we also focus on an 
observable that can be used to measure the kinetic energy 
of the system, which is given by,
\begin{eqnarray}
{O}&=&-\sum_{j=1}^{L-1}({c}^{\dag}_j{c}_{j+1}+\text{H.c.}). \label{obs}
\end{eqnarray}
The observable $O$ can be diagonalized going into the momentum basis $|k\rangle$,
and can be written as $O=\sum_k O_{kk}|k\rangle\langle k|$, 
where $O_{kk}$ are diagonal matrix elements of the operator $O$. We study the observational entropy using exactly the similar protocols described earlier but now replacing the 
real space basis $|x\rangle$ by the momentum basis $|k\rangle$ (in which the observable $O$ is diagonal).


\section{Results}
 Here we study the change of  observational entropy for the eigenstates of the  model Hamiltonian (\ref{nonint_model}) and during its  dynamical evolution. 
\subsection{Kinematics}

\begin{figure}
    \centering
    \includegraphics[width=0.48\textwidth, height=0.3\textwidth ]{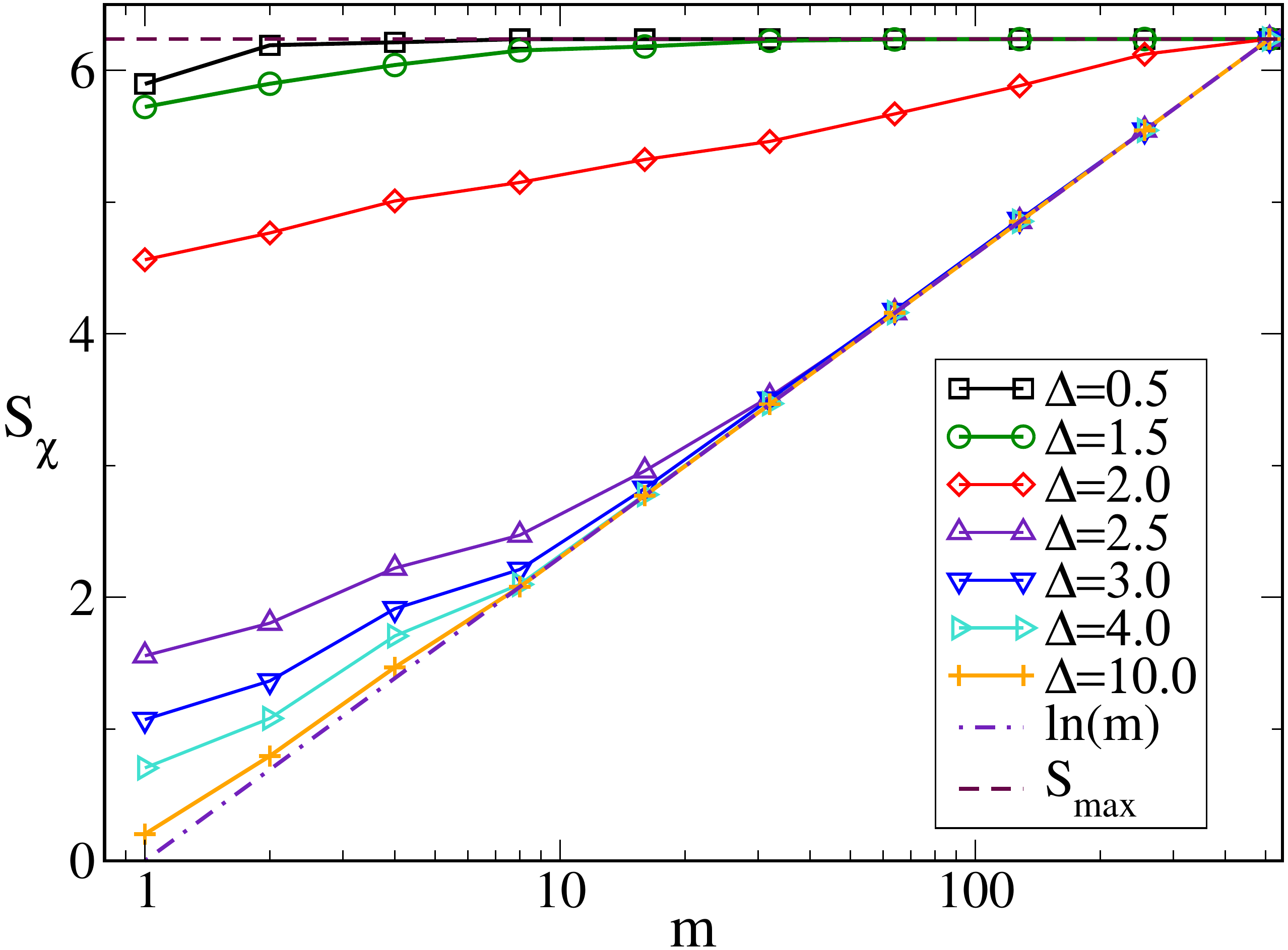}
    \caption{Variation of the observation entropy  $S_\chi$ with coarse-graining length $m$ the middle spectrum states  of the Hamiltonian $H$ for different values of $\Delta$.}
    \label{fig1}
\end{figure}


First we focus on the eigenstates of the the Hamiltonian ~(\ref{nonint_model}). For $\Delta=0$, the Hamiltonian ${H}$
can be easily diagonalized going into the momentum basis, i.e. 
${H}(\Delta=0)=-2\sum_k \cos k {n}_k$, where $k=2n\pi/L$
with $n=0,1 \cdot \cdot L-1$. The single-particle eigenvectors read as, 
\begin{equation}
    |\psi_k(\Delta=0)\rangle=\frac{1}{\sqrt{L}}\sum_{j=1}^{L} e^{ikj}{c}^{\dag}_j|0\rangle. 
\end{equation}
We consider a pure state and  ${\rho}=|\psi_k(\Delta=0)\rangle\langle\psi_k(\Delta=0)|$.
In case of the finest coarse-graining,  for which $V_{x_1} =1$, and for the  ground state ($k=0$) $p_{x_1}=1/L$ for all $x_1$, which  implies the observational entropy be simply equal to its Shannon entropy and  $S_\chi =\ln L$. On the other hand, for the  roughest coarse gaining  $\chi_I\equiv x_L$, $S_\chi=\ln L$. By referring to the property represented by Eq. (\ref{eq:fineC}),  for the ground state of the Hamiltonian 
${H}(\Delta=0)$ , for any coarse gaining $S_\chi = \ln L$, maximal  (see Fig.~(\ref{fig1})). 
On the other hand, in the limit $\Delta\to \infty$, the eigenstates of 
the Hamiltonian ${H}$ are completely localized, they read as,
\begin{equation}
    |\psi_{j_0}(\Delta\to\infty)\rangle={c}^{\dag}_{j_0}|0\rangle, 
\end{equation}
where $j_0$ is the site index. Hence, $p_{x_m}=1$ only for a particular $x_m$ for which the site $j_0 \in x_m$ and it is $0$ otherwise. It automatically implies, $S_\chi=\ln V_{x_m}$.


The change of the observational entropy with the coarse-graining length for the  middle of the spectrum states are shown in the  lower panel of Fig.~(\ref{fig1}) (we average over $10$ eigenstates from the middle of the spectrum around the energy $E\simeq0$). For the delocalized phase, 
$\Delta < 2$, observational entropy $S_\chi$ increases with the coarse-graining length and then it saturates to $S_{\text{max}}=\ln L$ for $m<L$.
On the other hand, in the localized phase, $\Delta > 2$, we find that beyond a certain coarse graning length $V_{x_m}=m > l_c$ ($l_c$ is some critical length scale depends on $\Delta$), $S_\chi =\ln V_{\chi}$.  

In the localized phase, given the single particle wave function has the following form, $\psi(j)\simeq e^{-|j-j_0|/\xi}$ ($\xi$ is the localization length and $j_0$ is site at which the wave function has a peak), 
one expects that for coarse-graining length $m > 2\xi$,  $S_\chi=\ln V_{x_m} $. These observations can be clearly  seen in figure.~(\ref{fig1}), where we find as $\Delta$ increases (localization length of the wavefunctions decrease), even for small values of coarse-graining length $S_{\chi}$ starts showing $\ln V_{x_m}$ scaling. However, it is worth 
pointing out that the growth of the $S_{\chi}$ as a function 
of $m$  in the delocalized phase much more rapid for the middle spectrum states compared to the ground state. While 
for the middle spectrum states $S_{\chi}=S_{\text{max}}=\ln L$
for $m<<L$, the ground state shows a slow logarithmic growth $S_{\chi}$ with coarse-graining size. This result is 
not very surprising given that typically one expects the degree of the delocalization (e.g. Shannon entropy) for high energy eigenstates to be much higher compared to the low energy ground state. Presumably, this fact  gets reflected in the results.

\subsection{Dynamics}

\begin{figure}
    \centering
    \includegraphics[width=0.48\textwidth,height=0.25\textwidth]{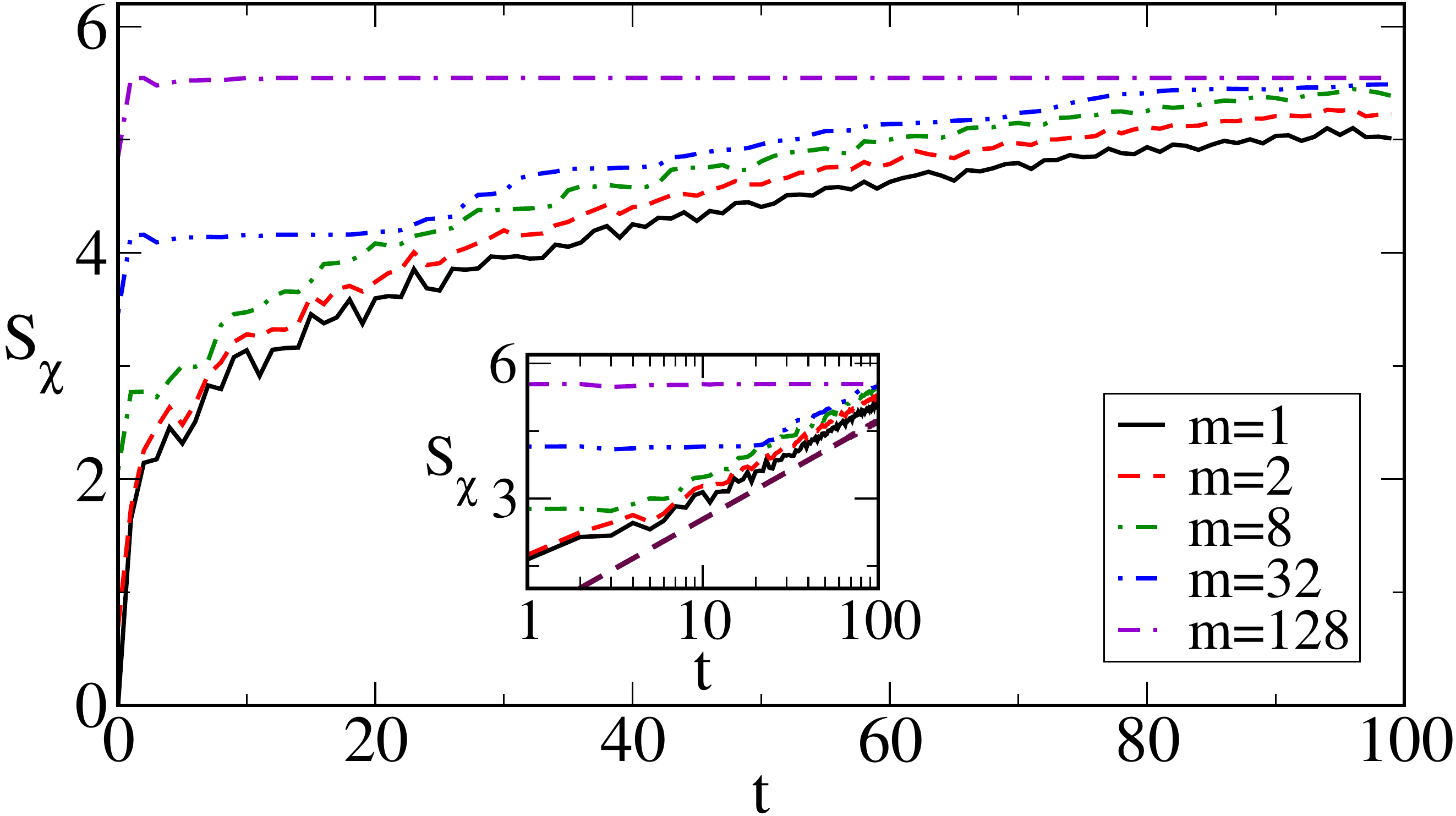}
        \includegraphics[width=0.48\textwidth, height=0.25\textwidth]{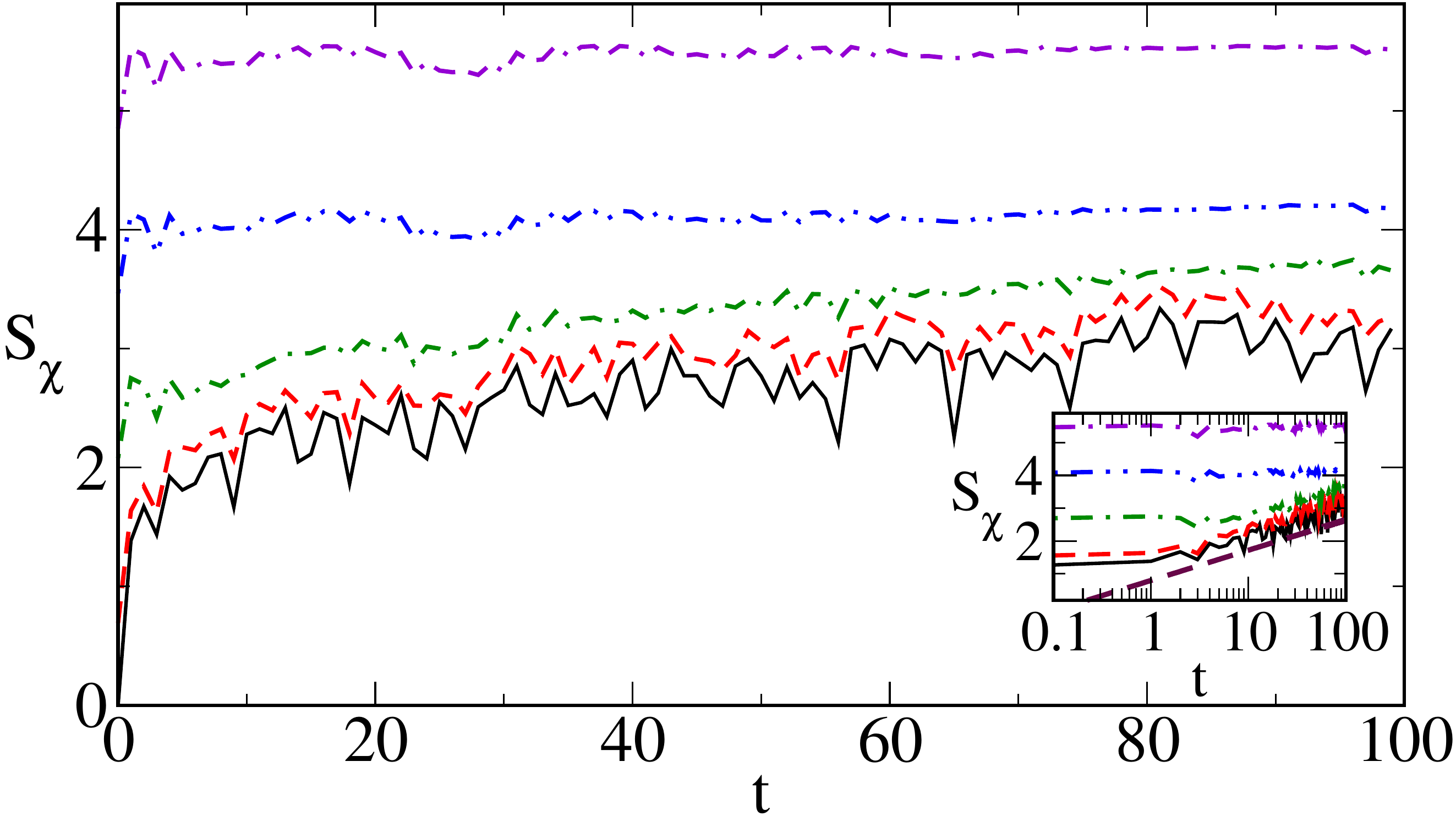}

    \includegraphics[width=0.48\textwidth, height=0.25\textwidth]{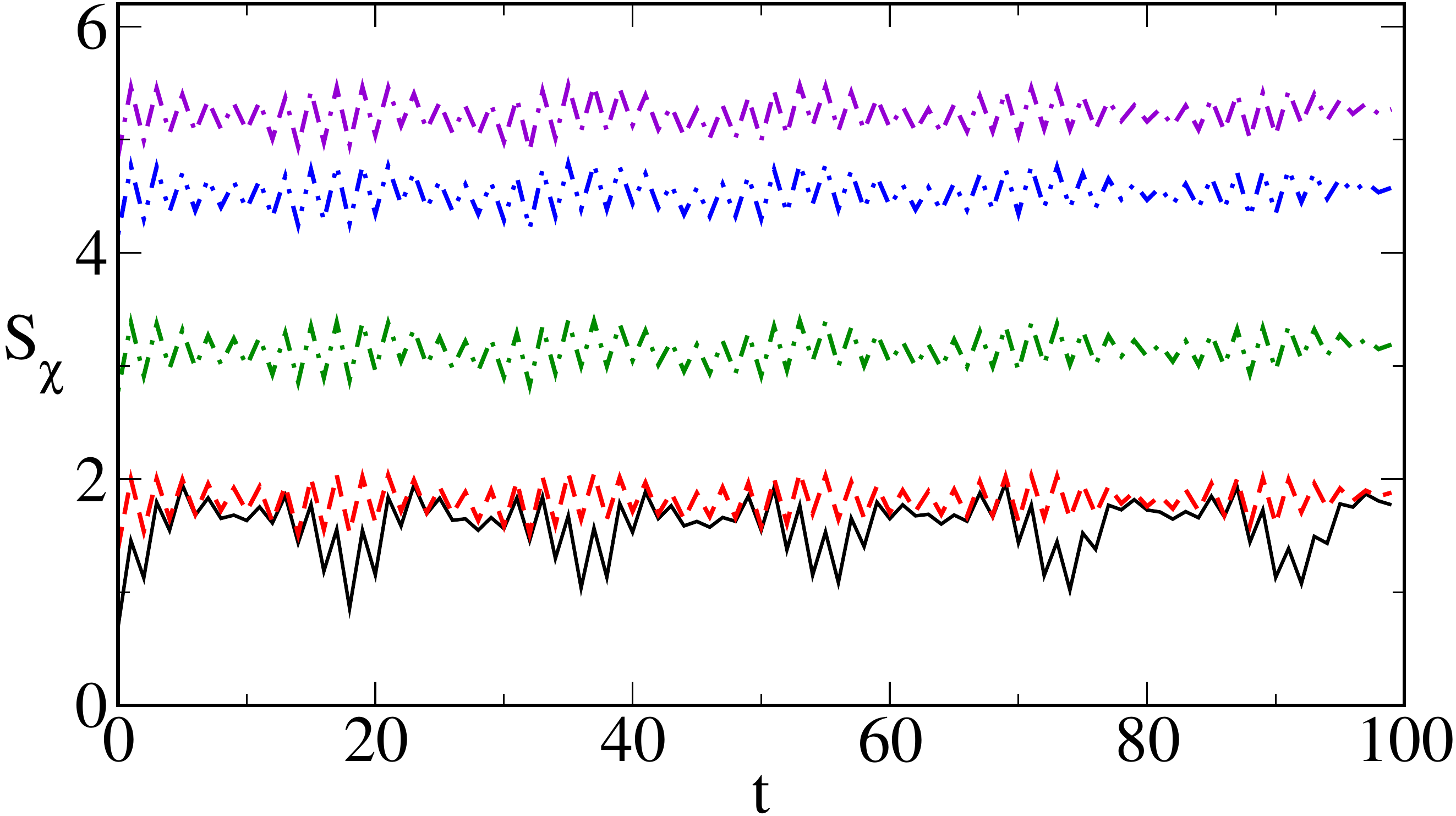}

    \caption{Variation of the observation entropy $S_\chi$ with time $t$ at the delocalized phase  $\Delta=1$ (upper panel),    transition point $\Delta=2$ (middle panel), and the localized phase $\Delta=3$ (lower panel) for different 
    coarse-graining. Insets in the upper and middle panel shows logarithmic growth $S_\chi$ with $t$ for $m<<L$. }
    \label{fig3}
\end{figure}

Here we study the behaviour of the observational entropy followed by the time evolution under the Hamiltonian $H$. 
We prepare the initial state such a way that electron is localized exactly on the middle of the lattice i.e., $|\psi(t=0)\rangle =|0_1 0_2 \cdot \cdot 1_{L/2} 0_{L/2+1}...0_L\rangle$. The time-evolved state reads 
as, 
\begin{equation}
    |\psi(t)\rangle=e^{-iHt}|\psi (t=0)\rangle.  
\end{equation}
Given for $\Delta=0$ the Hamiltonian can be easily diagonalized in the momentum basis, and 
the time-evolved state at a given time can expressed as, 
\begin{equation}
    \psi(t,j)=\langle j|\psi(t)\rangle=\frac{1}{L}\sum_k e^{ikj+2it\cos k}.
\end{equation}
In the thermodynamic limit, the series summation can be replaced by the integration. 
It is straight forward to check that observation entropy for the finest coarse-graining, i.e., $V_{x_1}=1$ will 
read as, 
\begin{equation}
    S_{\chi}=-2\sum_j(J_{j-L/2}(2t))^2\ln[J_{j-L/2}(2t)],
    \label{eq1}
\end{equation}
where $J_k()$ stands for Bessel function of 1st kind. 

Numerically, in the thermodynamic limit (in  Eqn.~(\ref{eq1}), by considering the   summation over sufficiently large numbers of  lattice sizes e.g. 256 or 512) the observational entropy in  Eqn.~(\ref{eq1}) is  $S_\chi \sim \ln t+\text{const}$. Hence, we expect to see $S_\chi \sim \ln t$ scaling for the systems we have studied here. 
Figure.~(\ref{fig3}) shows the variation of the observation entropy with time for different coarse-graining. 
In the delocalized phase,  the electron which was initially localized in the middle of the lattice, will spread over both sides of the lattice, i.e.,  the probability of finding that electron on the other sites away from the middle of the lattice  will also grow with time. It  implies that the 
observational entropy $S_\chi$ will grow with time. That growth should be very apparent when $V_{x_m}<<L$. 
The upper panel of Fig.~(\ref{fig3}) shows that indeed for $V_{x_m}<<L$, the growth of $S_\chi\sim\ln t$ (see inset of Fig.~(\ref{fig3}) (upper panel)). This is precisely 
what we obtained analytically for $\Delta=0$ as well. 
Interestingly, even in the transition point i.e. $\Delta=2$, we find a similar logarithmic growth (with lower slope) of the observational entropy with time $S_\chi \sim 0.45 \ln t$ (see the middle panel of Fig.~\ref{fig3}). However, the slope is smaller than 1 what we found in the delocalized phase. This could be a consequence of the anomalous transport which has been found in this model at the transition point~\cite{archak.2018}. 

On the other hand, in the localized phase, as one can expect, the observation entropy does not show any growth, it shows oscillatory behavior (see lower panel of Fig.~\ref{fig3} for $\Delta=3$ results). Also, note that for  the roughest graining  i.e. $V_{x_m}=L$, $S_\chi (t)=\ln L$.  Also, we find that for all time $S_{\chi}(t) \geq S_{\chi}(t=0)$ and  $S_{\chi_1}(t)> S_{\chi_2}(t)$ where $\chi_1 \huk \chi_2$.

\begin{figure}
    \centering
        \includegraphics[width=0.48\textwidth, height=0.3\textwidth]{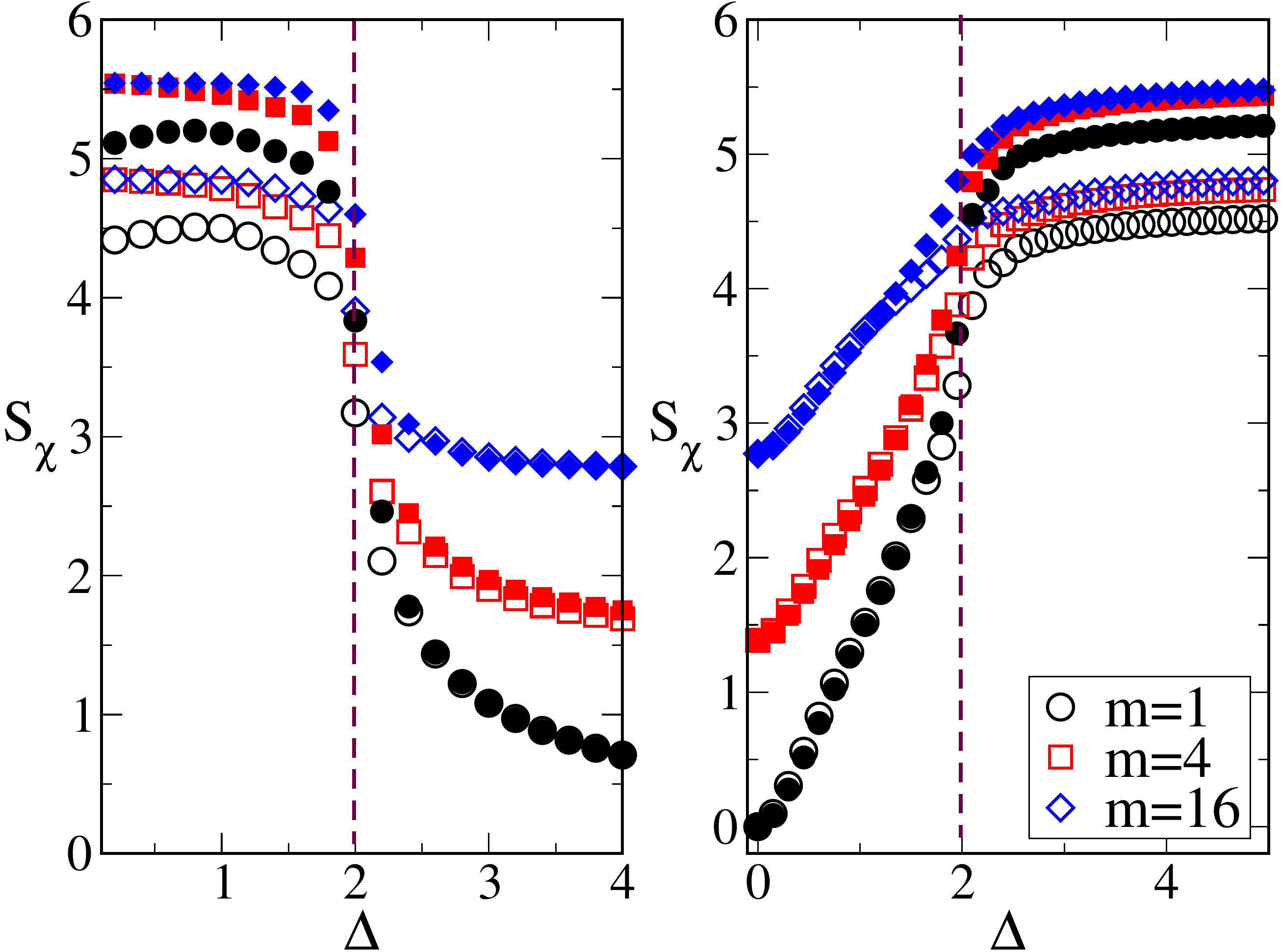}
       \caption{(Left panel)Variation of $S_\chi$ for the middle spectrum states  of the Hamiltonian $H$ with $\Delta$ for different coarse-graining length $m$. (Right panel)
    Variation of $S_\chi$ for the middle spectrum states  of the Hamiltonian $H$ with $\Delta$ for different momentum space coarse-graining length $m$. Results are for $L=128$ (inside empty symbols)   and $L=256$ (inside filled symbols).}
    \label{fig4}
\end{figure}

Next, we study the observation entropy as a function of $\Delta$ for different coarse-graining  for different system size is shown in Fig.~(\ref{fig4} (upper panel)).   In the delocalized phase, $p_{x_m}\simeq V_{x_m}/L$, which implies   $S_{\chi}\simeq \ln L$, hence $S_\chi$ increases with $L$. On the other hand, in the localized phase $S_{\chi}\simeq \ln V_{x_m}$, and does not depend on the system size as long as $V_{x_m}<L$. We also 
 repeat the calculations for the momentum space coarse-graining, because of the self-dual property of the AA model one expects to see $S_{\chi}$ does not depend on $L$ for $\Delta < 2$ (signature of localization) and $S_{\chi}$ increases with $L$ (signature of delocalization)  for $\Delta > 2$ (exactly opposite to what we had found in Fig.~\ref{fig4} (upper panel) for the real space coarse-graining). This is precisely what we observed in Fig.~\ref{fig4} (lower panel).

\begin{figure}
    \centering
        \includegraphics[width=0.48\textwidth, height=0.35\textwidth]{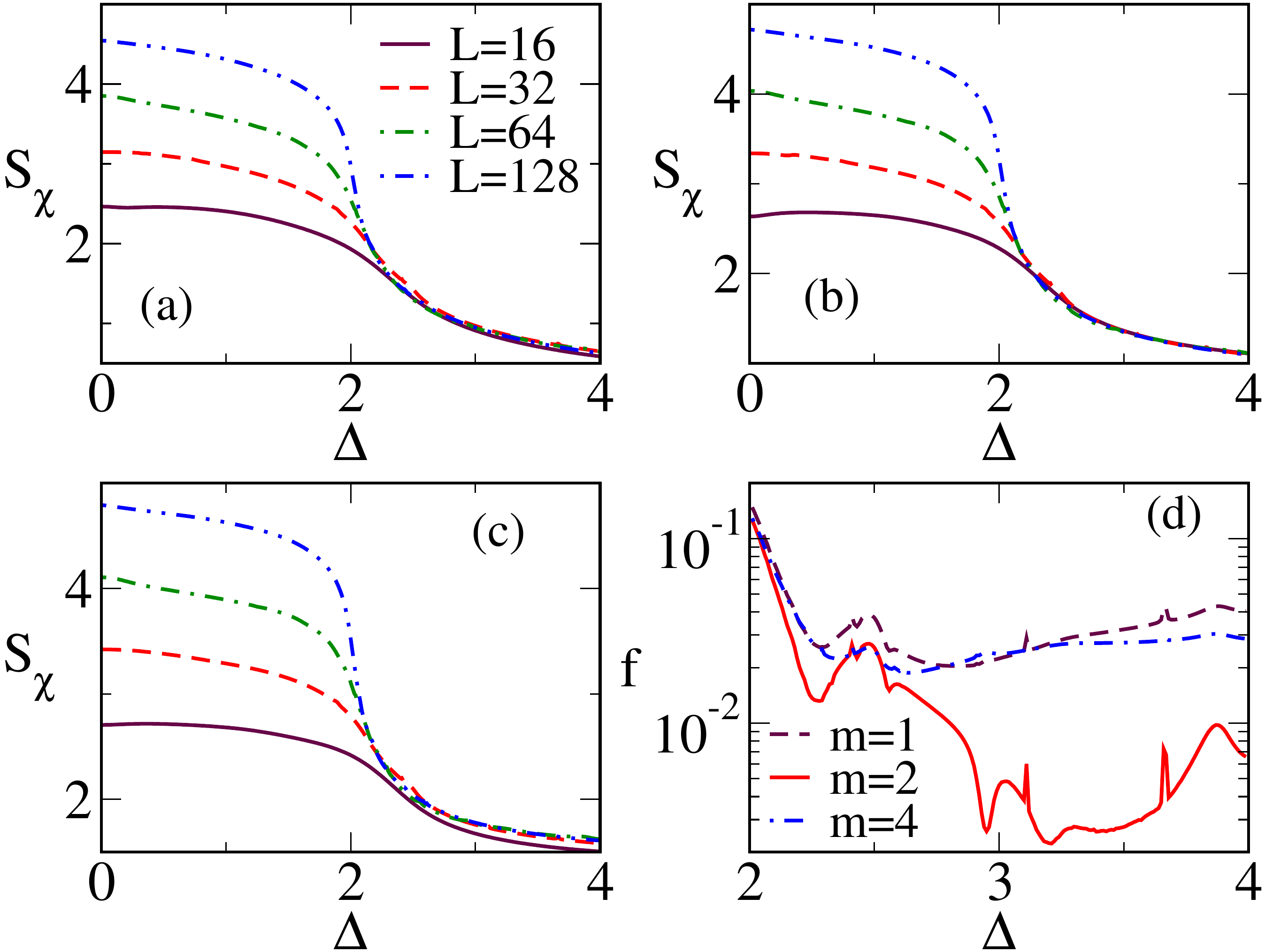}
       \caption{Variation of $S_\chi$ for the eigenstates  of the Hamiltonian $H$ with $\Delta$ for coarse-graining (real space) length  (a) $m=1$, (b) $m=2$, and (c) $m=3$ for $L=16$, 32, 64, 128.
    (d) Shows the variation of the fluctuation $f$ of $S_\chi$ with $\Delta$.}
    \label{fig5}
\end{figure}

Figure.~\ref{fig4} clearly demonstrates that the observation entropy shows data collapse ( for different values of $L$)
in the localized phase for any value of $m$ (as long as $m$ is smaller than the roughest coarse-graining length). However, for the finite size system, the question arises that how good is the data collapse. In contrast to the other diagnostic tools (e.g. Shannon entropy), observational entropy possesses an extra degree of freedom: the coarse-graining length.
Hence, we calculate the normalized fluctuation, which is defined as, 
\begin{equation}
 f(\Delta,m)=\frac{\sqrt{\langle S_{\chi}^2 \rangle -\langle S_{\chi} \rangle^2}}{\langle S_{\chi} \rangle}. 
\end{equation}
Here, $\langle \rangle$ stands for the average over different $L$ results. For the perfect data collapse, $f$ should be zero. On the other hand, the smaller the value of $f$, the better the data collapse. Figure.~\ref{fig5} shows the variation of $S_{\chi}$ with $\Delta$ for $m=1$, 2, 4. Looking at the figures, it is difficult to distinguish which coarse-graining length corresponds to better data collapse. Hence, we show the variation of $f$ with $\Delta$ in the localized region ($\Delta> 2$). We find that $f$ is the minimum for $m=2$, which makes $m=2$ the optimal coarse-graining length for our case.

\section{Conclusions\label{sec:conclu}}

    Here we have studied the coarse-grained 
  observational entropy for the 1D AA model, which supports localization-delocalization transition.
  First, we find that for the middle of spectrum states 
  as one increases the coarse-grain size, the observational 
  entropy very rapidly saturates to the upper bound i.e. $\ln L$. On the other hand, in the localized phase, typically if the coarse-grain size is greater than  the order of localization length, the observational entropy increases as logarithmic of the coarse-grain size. Also, if the coarse-grain size is less than the system size, the observation entropy in the delocalized phase scales logarithmically with system size in contrast to the localized phase, where it does not depends on system size.
  
  We have also investigated the fate of the observational entropy followed by a quantum quench under the AA Hamiltonian, where the particle was initially localized in the centre of the lattice. We  found for a given coarse-graing (as long as the coarse-grain size is much smaller than the system time) in the delocalized phase (as well as in the localization-delocalization transition point), it grows with time logarithmically, while in the localized phase it does not grow at all. Finally, we also show that the observational entropy corresponding to the momentum space coarse-graining can clearly manifest the self-dual property of the AA model.   We also have explicitly demonstrated that one can find an optimal coarse-graining length for which the finite size
scaling shows much better data collapse.

  While in this work, we had restricted our-self to the non-interacting systems, in the future  work it will be interesting to investigate 
  the behaviour of the observational entropy in the context of many-body localized systems, where the localization of many-body wave function takes place in the 
  many-body fock space \cite{smukerjee.2018,smukerjee.2022}. As observational entropy is a directly measurable quantity, we would like to investigate the studies in the current paper in QC platforms.

\section{Acknowledgements}
RM acknowledges the DST-Inspire fellowship by the Department of Science and Technology, Government of India, SERB start-up grant (SRG/2021/002152). SA acknowledges the start-up research grant from
SERB, Department of Science and Technology, Govt. of India (SRG/2022/000467).  Authors  thanks Vinod Rao, Ranjith V and Shaon Sahoo for fruitful discussions.  
\bibliography{thermo}
\end{document}